\documentclass[12pt]{article}
\usepackage{pdfpages}
 \pdfoutput=1

\makeatletter
\@addtoreset{equation}{section}
\makeatother

\title{Optimal Boundary Kernels
and Weightings for Local Polynomial Regression
}

\author{
Alexander Sidorenko \\
and \\
Kurt S. Riedel \\%\footnote{riedel@mfdd4.nyu.edu}
Courant Institute of Mathematical Sciences \\
New York University \\
New York, New York 10012-1185}

\date{1993}

\begin{document}

\maketitle

\begin{abstract}
Kernel smoothers are considered near the boundary of the interval.
Kernels which minimize the expected mean square error are derived.
These kernels are equivalent to using a linear weighting function
in the local polynomial regression.
%Our kernels are more robust
%with respect to the incorrect choice of the halfwidth
%than other standard estimates.
It is shown that any kernel estimator that satisfies the moment conditions up to 
order $m$ is equivalent to a local polynomial regression of order $m$ with
some non-negative weight function if and only if the kernel has at most $m$ sign changes.
A fast algorithm is proposed for computing the kernel estimate in the boundary region
for an arbitrary placement of data points.
\end{abstract}

\newpage

\section{Introduction}\label{I}

We consider a traditional nonparametric curve estimation problem.
Noisy measurements of an unknown function $f$ are given:
$y_i=f(x_i)+\varepsilon_i$,
$i=1,2,\ldots ,N$,
where the errors, $\varepsilon_i$, are independent random variables
with zero mean and variance equal to $\sigma^2$.
The function, $f$, is assumed to have $p$ continuous derivatives.
Our goal is to estimate its $q$th derivative ($0\leq q<p$)
in the interval of measurements.

Presently, three methods are widely used:
kernel smoothers, local polynomial regression,
and smoothing splines.
Excellent reviews of kernel smoothing can be found in the monographs by
Hardle (1990) and M\"uller (1988).
%Rosenblatt (1956), Parzen(1962), Pristley and Chao (1972), Singh(1979),
%Gasser and M\"uller (1979,1984),
%Gasser, M\"uller and Mammitzsch (1985), M\"uller (1984,1988)
In this method, the estimator of $f^{(q)}$ has the form
$
 \widehat{f^{(q)}}(t)
  =
 \sum_{i=1}^N K(t,x_i) y_i
  .
$
To ensure consistency,
the kernel must satisfy certain moment conditions.
The local polynomial regression method is described in works by
Cleveland (1979), Lejeune (1985),
Fan and Gijbels (1992),
Fan (1993), Hastie and Loader (1993, including comments by
Fan and by Marron, and by M\"uller),
Jones (1994).
For a given estimation point $t$, a fitting polynomial
$\sum_{j=0}^{p-1} a_j x^j$ is sought to minimize
$
   \sum_{i=1}^N w_i(t)
            \left(\sum_{j=0}^{p-1} a_j(x_i-t)^j - y_i\right)^2 .
$
Then $q!a_q$ is taken as the estimate of $f^{(q)}(t)$.
Parameters of the method are the non-negative weights, $w_i$,
and the order of polynomial fitting, $p-1$.
The weights are usually scaled as
$w_i(t) = W \left(\frac{x_i-t}{h}\right)$
where $W$ is a non-negative function on [-1,1].

Silverman (1984) showed that smoothing splines
are equivalent to a special case of kernel smoothers.
Similarly,
the local polynomial regression is equivalent to
a kernel smoother $K(t,x_i)=w_i(t)P(x_i-t)$
where $P(x)$ is a polynomial of order $p-1$ in $x$
whose coefficients may depend on $t$
(see M\"uller (1987), Jones (1994)).
We show that {\em any kernel estimator
that satisfies the moment conditions up to order $p-1$
is equivalent to a local polynomial regression of order $p-1$
%with some non-negative weight function
if and only if the kernel has at most $p-1$ sign changes}.

The optimal kernel support
to minimize the  mean square error (MSE)
for a prescribed kernel shape dates back to
Grenander and Rosenblatt (1957).
%In the case of interior kernels,
Away from the beginning and end of the data,
the optimal kernel shape (which minimizes the MSE)
was determined by
Gasser and M\"uller (1979), M\"uller (1984, 1988),
Gasser, M\"uller and Mammitzsch (1985).
The bias of a nonparametric smoothing estimator
typically is much higher in the boundary region than in the interior
(see Rice and Rosenblatt (1981)).
Thus, minimization of MSE in the boundary region is important
and is a central focus of this article.
We also notice that estimation near the boundary
includes forecasting as a special case when $t<0$ or $t>1$.

Near the boundary of the observation interval,
the kernel shape has to be smoothly transformed to allow
nonsymmetrical support.
Subject to the moment conditions and the smooth boundary transformation,
we choose the kernel shape
to minimize MSE.
Various boundary transformation techniques
are discussed in literature (Jones (1994)).
In the generalized jackknifing boundary correction,
Aitken extrapolation is used to reduce the order of the bias
(Rice (1984)).
M\"uller (1994) and Jones (1994)
propose general methods which provide
boundary transformation of the prescribed smoothness
for any interior kernel.
All these transformations satisfy the moment conditions
and provide smooth estimates.
They do not, however, minimize MSE near the boundary.

The boundary problem is solved automatically in
the local polynomial regression method:
the boundary region points are treated similarly
to the interior points.
In terms of equivalent kernels, the factor
polynomial $P$ depends on the estimation point
and changes smoothly near the boundary.
This produces a smooth boundary transformation
of the equivalent kernel.
The weighting function, however,
which minimizes MSE in the interior,
does not necessarily minimize MSE in the boundary region.

{\em
In the present work, we investigate the problem of choosing
a boundary kernel (or equivalently, selecting a weight function
for the local polynomial regression)
which minimizes MSE
while providing continuity (smoothness) of the estimate.
}
For a given kernel halfwidth in the boundary region,
we derive the optimal boundary kernel
and its approximation for large $N$, the asymptotically optimal kernel.
Our boundary kernels have a simple form and are polynomials
of the same order as the optimal interior kernels.
We show that in the case when the kernel halfwidth is constant
in the boundary region, the estimate can be computed
in ${\cal O}(N_T+N_E)$ operations,
where $N_T$ is the number of data points within the support,
and $N_E$ is the number of boundary region points
where $f^{(q)}$ is to be estimated.
The placement of data points can be arbitrary.

In Section~\ref{II}, we review MSE of kernel smoothers.
%Our formulation allows asymmetric support
%in order to include the boundary case.
In Sections~\ref{IV} and~\ref{OK},
we derive the optimal boundary kernel
which minimizes the leading order MSE.
In Section~\ref{V}, we concentrate on the limiting case
where the data points are spaced approximately regularly
as $N\rightarrow\infty$ .
We derive simple analytical expressions for the optimal boundary kernels.
%with $p=q+2$.
In Section~\ref{LPR1}, we investigate the equivalence
of the kernel smoother estimators and the local polynomial regression.
In Section~\ref{LPR2}, we show that
the optimal weighting in the boundary region
is a linear function.
In Section~\ref{LPR3}, we discuss estimation for non-equispaced data.
In Section~\ref{Comp}, we compare MSE of various estimators
and show that the asymptotically optimal kernel is robust against
the misspecification of the kernel halfwidth.

\section{Expected Mean Square Error of Kernel Smoothers}\label{II}

Let $f(t)$ have $p$ continuous derivatives and assume that
$f^{(p)}\neq 0$ in the domain. 
We consider kernel estimators of $f^{(q)}(t)$ ($q<p$) of the form:
\begin{equation}\label{2.1}
\widehat{f^{(q)}}(t)  = \sum_{i=N_L}^{N_R}  K(t,x_i) y_i
\ .
\end{equation}
The left and  right endpoints of the summation are free parameters which
we optimize. We denote $x_{N_L}$ by $x_L$, $x_{N_R}$ by $x_R$, and define
$h_L \equiv t -x_L$ and  $h_R \equiv x_R -t$.
We define  the kernel halfwidth, $h \equiv {h_L + h_R \over 2}$
and $N_T = N_R-N_L +1$.
%We note that $N_T \sim 2Nh$ for equispaced data.
%We define the $N_T$--component vector,
%$\mu = N_T h^q(K(t,x_L), \ldots K(t,x_R))^T$,
%where we suppress the kernel dependencies on $t$, $N_L$ and $N_R$,
%and rescale the kernel such that $\mu$ is of order one.
For equispaced data,
away from the data boundary,
the most common kernel smoothers are
{\em
scale parameter kernels}:
$K(t,x_i) = \frac{1}{Nh^{q+1}} G\left(\frac{x_i-t}{h}\right)$.

For given values of $t$, $N_L$, and $N_R$, 
we say a kernel, $K(t,x_i)$,
with halfwidth, $h$,
is of type $(q,p)$ if
it satisfies the {\em moment conditions}:
%\begin{equation}\label{2.2}
\begin{eqnarray}\label{Mom}
 \frac{1}{Nh} \sum_{i=N_L}^{N_R}
    \frac{1}{m!} \left(\frac{x_i-t}{h}\right)^m
 \mu_i
 & = & \delta_{m,q} \ , \;\;\;\;\;\; m=0,\ldots ,p-1
\; ,
\\ \nonumber
\\ \nonumber
 B_p \; = \; \frac{1}{Nh} \sum_{i=N_L}^{N_R}
    \frac{1}{p!} \left(\frac{x_i-t}{h}\right)^p \mu_i
 & \neq & 0 \; ,
\end{eqnarray}
%\end{equation}
where
$\mu_i = N h^{q+1} K(t,x_i)$.
Kernels of type $(q,p)$ estimate the $q$th derivative
of the function with a bias error of order $O(h^{p-q})$.
The bias term can not be eliminated by setting $B_p=0$:
in such a case, the kernel would be of a higher order.

We assume that $h_L $  and  $h_R $  are small ($N_L <<N $ and  $N_R <<N $),
%relative to the characteristic scalelength of} $f(t)$, and therefore
and expand $f(x_i)$ in a Taylor series about $f(t)$.
For a kernel smoother
of type $(q,p)$, the bias of the estimator is
\begin{equation}\label{2.4}
 E\left[ \widehat{f^{(q)}} (t)\right]-  f^{(q)}(t) =
B_p f^{(p)}(t) h^{p-q}
+O(h^{p-q+1})
.
\end{equation}
The variance of the kernel estimator is
\[ %\begin{equation}\label{2.3}
{\rm Var} \ [ \widehat{f^{(q)}}(t)] =
 \frac{\sigma^2}{N^2 h^{2q+2}}
\sum_{i=N_L}^{N_R} \mu_i^2
\  \ .
\] %\end{equation}
We define
$m_2(\mu)=\frac{1}{Nh}\sum_{i=N_L}^{N_R} \mu_i^2 $.
In the case of
scale parameter kernels,
$m_2(\mu)$ converges to $\int_{-1}^1 G(y)^2dy$.
Correspondingly, $B_p$ converges to
$\int_{-1}^1 G(y)\frac{y^p}{p!}dy$.
Thus the leading order MSE is
\begin{equation}\label{2.5}
R(t) =
B_p^2 |f^{(p)}(t)|^2  h^{2(p-q)}
+ {\sigma^2 m_{2}(\mu) \over N h^{2q+1}} \
,
\end{equation}
where the corrections are $O(h^{2(p-q)+1})$.
%In Equation (\ref{2.5}), we have replaced $N_T$ by $2Nh$.

Solving (\ref{2.5}) for the optimal value
of the kernel scale size, $h_0$, yields
\begin{equation}\label{2.6}
 h_0 \; = \;
 \left( \frac{2q+1}{2(p-q)}
        \frac{\sigma^2 m_{2}(\mu)}{B_p^2 N |f^{(p)}(t)|^2}
     \right)^{\frac{1}{2p+1}}
  \; .
\end{equation}
The ratio
$N|f^{(p)}|^2 / \sigma^2$
is the expansion parameter for choosing the kernel's halfwidth.
%If the ratio is less than of order 1,
%there is no satisfactory choice of a local support,
%and we should choose the whole interval $[0,1]$ as a support for the kernel.
This ratio can be normalized to one by
changing variables to
$\tilde{t}=\alpha t$ where
$\alpha =\left|\frac{N|f^{(p)}|^2}{\sigma^2}\right|^{\frac{1}{2p+1}}$.
This transformation maps the interval $[0,1]$ to $[0,\alpha ]$.
Then for each specific point $t\in [0,\alpha ]$, we consider
a subinterval of length 1
(which contains about $N/\alpha$ measurements)
around this point.

For the choice of kernel halfwidth given by (\ref{2.6}),
%the total square error of (\ref{2.5}) is
the leading order MSE equals to
\begin{equation}\label{2.7}
R(t) \; = \;
K_{q,p}  \left| B_p f^{(p)}(t)\right|^{\frac{4q+2}{2p+1}}
\left|{ \sigma^2 m_{2}(\mu)
\over N }
\right|^{\frac{2(p-q)}{2p+1}}
% B_p^2 |f^{(p)}(t)|^2 }
%({ N} \over N)^{2p}
\ ,
\end{equation}
where $K_{q,p}\equiv \left| {2q+1 \over 2(p-q)}\right|^{2(p-q)\over 2p+1} +
\left| {2(p-q) \over 2q+1}\right|^{2q+1\over 2p+1}$.
The optimal $h$ is proportional to $N^{-\frac{1}{2p+1}}$,
and MSE
is proportional to $N^{-\frac{2(p-q)}{2p+1}}$.
If  $f(t)$ has $\tilde{p}$  continuous derivatives, where
$q \le \tilde{p} \le p$, the optimal halfwidth scales as
$N^{-\frac{1}{2\tilde{p}+1}}$,
and the total square error
is proportional to $N^{-\frac{2(\tilde{p}-q)}{2\tilde{p}+1}}$.
The convergence rate given by (\ref{2.7}) is theoretically optimal for
functions with precisely ${p}$ continuous
derivatives (Stone (1982)). 
The leading constant depends on the kernel shape, however.

%\section{Smoothness of the Kernel Estimate}\label{III}

The estimate, $\widehat{f^{(q)}}(t)$ in (\ref{2.1}),
is $C_l$ if and only if
$K(t,x_i)$ is $C_l$ over [0,1] for every $x_i$.
To impose smoothness of order $l$ on $\widehat{f^{(q)}}$,
some researchers require that the kernel
and its first $l$ derivatives
vanishes at the boundaries of the support.
{\em
This is not necessary when the boundary
does not change as $t$ varies:
in particular, when the support borders
one of the ends of the interval $[0,1]$.
}
In the present paper,
we consider the case $l=0$,
so we seek an estimate, $\widehat{f^{(q)}}(t)$,
that is continuous in $t$.

\section{Orthogonal Polynomial Representation of the Optimal Kernels}\label{IV}

In view of moment conditions (\ref{Mom}),
we expand the kernel in orthogonal polynomials.
Let $P_k$ be a polynomial of order $k$ $(k=0,1,\ldots )$
such that
$\frac{1}{Nh} \sum_{i=N_L}^{N_R}
  P_k\left(\frac{x_i-\overline{x}}{h}\right)
  P_j\left(\frac{x_i-\overline{x}}{h}\right)
   = g_k \delta_{kj}$
where $\overline{x}=\frac{1}{2}(x_L+x_R)$
and $g_k$ is a normalization.
We expand
$K(t,x)=\frac{2}{N h^{q+1}}
 \sum_k b_k P_k\left(\frac{x-\overline{x}}{h}\right)$.
The moment conditions are rewritten as
$\sum_k C_{kj} b_k = \delta_{qj}$ for $j=0,\ldots ,p-1$,
where
$C_{kj} = \sum_{i=N_L}^{N_R}
  P_k\left(\frac{x_i-\overline{x}}{h}\right)
 \frac{1}{j!} \left(\frac{x_i-t}{h}\right)^j$.
The matrix $C_{kj}$ is upper triangular,
and its diagonal entries are not zero.
We solve for $b_0, b_1 \ldots  b_{p-1}$:
\begin{eqnarray} \label{E4}
 b_j & = & 0 \;\;\;\;\;\; {\rm with} \;\; j=0,\ldots ,q-1 \; , \nonumber \\
 b_q & = & \frac{1}{C_{qq}} \; , \\
 b_j & = & - \frac{1}{C_{jj}}\sum_{i=q}^{j-1} C_{ij}b_i
                  \;\;\;\;\;\; {\rm with} \;\; j=q+1,\ldots ,p-1 \; .
            \nonumber
\end{eqnarray}
Any kernel satisfying the moment conditions has the coefficients
$b_0, b_1 \ldots  b_{p-1}$ prescribed by (\ref{E4})
while coefficients $b_p,b_{p+1},\ldots$ are free parameters.
The leading order bias equals
$f^{(p)}(t) h^{p-q} \sum_{k=q}^p C_{kp} b_k$.
The summation stops at $k=p$ because $C_{kp}=0$ for $k>p$.
In the absence of boundary conditions,
MSE attains the minimum when $b_k=0$ for $k>p$.
Then MSE is a quadratic function of $b_p$,
and the optimal value $b_p$ can be easily found:
\begin{equation}\label{4.2}
 R(t)  =  \frac{\sigma^2}{N h^{2q+1}}
 \sum_{k\geq q} g_k b_k^2   +
 \left( f^{(p)}(t) h^{p-q} \sum_{k=q}^p C_{kp} b_k\right)^2
 \; .
\end{equation}
If there are no boundary conditions, the minimum is attained when
$b_k=0$ for $k>p$ and
\begin{equation}\label{Bopt}
 b_p \; = \; - \left(\sum_{k=q}^{p-1} C_{kp} b_k\right) \left/
                    \left( C_{pp} + \frac{g_p \sigma^2}
                                   {C_{pp} N h^{2p+1} (f^{(p)}(t))^2}
                             \right) \right.
    \; .
\end{equation}
We name the kernel,
$K(t,x)=\frac{2}{N h^{q+1}}
 \sum_{k=q}^p b_k P_k\left(\frac{x-\overline{x}}{h}\right)$
with coefficients $b_k$ given by (\ref{E4}) and (\ref{Bopt}),
the {\em optimal kernel}.
When $t<0$ or $t>1$,
this optimal kernel can be used for forecasting.

If we optimize the kernel shape over the whole interval, [0,1],
the optimal kernel still minimizes the leading order MSE.
However, the bias estimate in (\ref{2.4}) is based on
the Taylor series expansion of $f(x_i)$ about $f(t)$.
When the kernel support is not small with respect to the characteristic
scale of the $f(t)$,
the actual bias error may be unrelated
to the leading order MSE in (\ref{4.2}).

\section{Optimal Boundary Kernels}\label{OK}

Away from the ends of the data,
we center the support of the kernel about the estimation point.
As the estimation point, $t$, approaches the left endpoint of
the interval $[0,1]$,
the kernel halfwidth, $h(t)$, eventually becomes equal to $t$,
and the support touches the left end, 0.
We name the point where it happens the {\em touch point}
and denote by $t_0$.
At this point, $h(t_0)=t_0$.
We refer to the subinterval $[0,t_0]$ as the
{\em left boundary region}.
When the estimation point belongs to this region,
we place the left end of the support
at the left end of the interval:
$t-h_L(t)=0$.
The right end of the support equals $2h(t)$,
where $h(t)=\left(h_L(t)+h_R(t)\right) /2$.

In the boundary region, we may consider two options:
fixed or variable kernel halfwidth.
Variable halfwidth choice, $h=h(t)$, allows to reduce MSE
by taking into account changes in $f^{(p)}$
(see M\"uller and Stadtm\"uller (1987)).
We show in Section~\ref{V} that using a fixed halfwidth
in the boundary region results in the minimum MSE as
$N\rightarrow\infty$.
Moreover, using a fixed halfwidth in the boundary region
has three important advantages:
{\em (i)}
prior estimation of $f^{(p)}(t)$ in the boundary region
is not required,
{\em (ii)}
the continuity of the estimate can be ensured
while using kernels which do not vanish at the ends of
the kernel support
(and thus MSE decreases),
and
{\em (iii)}
computational costs can be reduced drastically as we show below.

Indeed, if the support, $[0,2h]$, is the same for all estimation
points $t\in [0,h]$, then the same system of orthogonal polynomials
$P_k$ can be used for all $t$.  The only dependence of formulae
(\ref{E4}) on $t$ is due to the fact that $C_{kj}=C_{kj}(t)$ is a
polynomial of order $j$ in $t$.  The orthogonal polynomials
$P_0\left(\frac{x}{h}-1\right), P_1\left(\frac{x}{h}-1\right),\ldots ,
P_p\left(\frac{x}{h}-1\right)$ and the coefficients of the polynomials
$C_{kj}(t)$ can be calculated in ${\cal O}(N_T)$ operations where
$N_T$ is the number of data points in the boundary region.  We also
need ${\cal O}(N_T)$ operations to compute the inner products $ s_j =
\sum_{i=1}^{N_T} P_j\left(\frac{x_i}{h}-1\right) y_i $ for
$j=q,q+1,\ldots ,p$.  After that, for each $t$, we need only ${\cal
O}(1)$ operations to compute $b_q(t),b_{q+1}(t),\ldots ,b_p(t)$ and
the estimate, $ \widehat{f^{(q)}}(t) = \sum_{j=q}^p b_j(t) s_j .  $

In general, the same computational scheme also can be applied when
the kernel has a fixed support, is a polynomial of a given order,
and the coefficients of its orthogonal polynomial expansion
are rational functions in $t$.

%In practice, choosing the optimal halfwidth is very complicated,
%see Hall and Johnstone (1992).

\section{Optimal Boundary Kernels in Continuum Limit for $p=q+2$}\label{V}

Equispaced data
arise increasingly often with the widespread use of digital
signal processing.
In the limiting case,
when the data points are regularly spaced and their number is large,
analytic expressions for optimal interior kernels were obtained by
Gasser, M\"uller and Mammitzsch (1985).
In this section, we consider even more broadly defined limit where
every subinterval of length
$\frac{1}{\alpha}$ with $\alpha={\cal O}\left( N^{\frac{1}{2p+1}}\right)$
contains $\left(\frac{1}{\alpha}+o(1)\right)N$ data points
as $N\rightarrow\infty$.

In the continuum limit, the discrete kernel function,
$K_N(t,x_i)\;$, $i=1,2,\ldots ,N$,
is replaced with a function $K(t,x)\;$, $x\in [0,1]$.
Namely, if $x_i\rightarrow x$
then \mbox{$N\cdot K_N(t,x_i)\rightarrow K(t,x)$}.
%In approximating discrete sums by continuous integrals,
%we are making an error of $O\left(\frac{1}{N}\right)$.
%If the data points are equidistantly spaced,
%the error can be reduced to $O\left(\frac{1}{N^2}\right)$.
%
%In the continuum limit,
Summation over $x_i$ is replaced with integration over $x$,
and the discrete orthogonal polynomials
are replaced with their
continuous analog, Legendre polynomials (see Appendix~A).

The {\em optimal interior kernel} of type $(q,p)$
(see Granovsky and M\"uller (1989))
is the kernel which minimizes the leading order MSE
subject to the constraint that the number of sign changes
in the open interval of the kernel support is at most $p-2$.
The prescribed value of sign changes is the minimal possible,
see M\"uller (1985).
As we show in Section~\ref{LPR1}, the number of sign changes of the kernel
is related to the existence of an equivalent
estimator in the local polynomial regression.

From now to the end of this section,
we restrict ourselves to the case of $p = q+2$.
In this case,
the optimal interior kernel, in the continuum limit,
can be represented as
$K(t,x)=\frac{1}{h^{q+1}}G\left(\frac{x-t}{h}\right)$,
where
\begin{equation}\label{E119}
G(y) \; = \;
 \gamma_q \cdot (
 P_q(y)-P_{q+2}(y)
 )
\; ,
\end{equation}
$P_q, P_{q+2}$ are the Legendre polynomials,
and
$ \gamma_q = \frac{1}{2} \prod_{k=1}^{q}(2k+1) $.
The leading order MSE is minimal for the halfwidth
\begin{equation}\label{E9}
 h_0(t) \; = \;
 \left(\frac{4(2q+3)(2q+5)\sigma^2 \gamma_q^2}{Nf^{(p)}(t)^2}
 \right)^{ \frac{1}{2p+1}}
 \; .
\end{equation}
With this halfwidth, the leading order MSE equals
$
 \frac{2q+3}{2q+1}
 \frac{\sigma^2\gamma_q^2}{N h_0(t)^{2q+1}}
$.
The optimal interior kernel vanishes at the ends of the support
and thus ensures the continuity of the estimate.

We seek a boundary kernel in the form:
$
 K(t,x)=\frac{1}{ h ^{q+1}}
        G\left(\frac{t}{ h }-1,\frac{x}{ h }-1\right)
 .
$
The function $G(z,y)$ is
the {\em normalized boundary kernel},
and its domain is $y\in [-1,1]$, $z\in [-1,0]$
(we notice that $z\leq 0$
because $h_L < h_R$ in the left boundary region).
We may use the same normalized kernel, $G$, to represent a boundary kernel
in the {\em right boundary region} (where $t$ is close to 1) :
$
 K(t,x)=\frac{(-1)^q}{ h ^{q+1}}
        G\left(\frac{1-t}{ h }-1,\frac{1-x}{ h }-1\right)
$
with $x\in [1-2 h ,1]$.

Using the Legendre polynomials, $P_j$,
we expand the normalized boundary kernel:
$
 G(z,y) = \gamma_q \sum_j b_j(z) P_j(y)
 .
$
From Appendix~A, equation (\ref{E4}) reduces in the continuous limit to
\begin{equation}\label{E25}
 b_j = 0 \;\;\;\; {\rm with} \;\; j=0,\ldots ,q-1 \; ;
     \;\;\;\;\;\;\;\;
 b_q = 1 \; ;
     \;\;\;\;\;\;\;\;
 b_{q+1} = (2q+3) z \; .
\end{equation}
These are an equivalent of the {\em moment conditions} for the kernel.
Correspondingly,
the leading order MSE of the kernel estimator
in Eq.~(\ref{4.2}) reduces to
\[ R(t) \; = \;
 \frac{\sigma^2\gamma_q^2}{N h_0(t)^{2q+1}}
\overline{R}
 \left(\frac{t}{ h }-1\right)
%\left(\frac{t}{h}-1,\frac{h}{h_0(t)},\frac{b_q}{\gamma_q}\right)
  \; , \]
where
$\overline{R}$ is
the {\em normalized risk}:
\begin{eqnarray}\label{E14}
  \overline{R}
   (z)
 & = & \frac{2}{\beta^{2q+1}}
   \left( \frac{1}{2q+1} + (2q+3)z^2
      + \sum_{j\geq q+2} \frac{b_j^2}{2j+1} \right)
        \nonumber  \\
         & + &
          (2q+3)(2q+5)
      \beta^4
   \left( \frac{1}{2q+3} - z^2
    + \frac{2b_{q+2}}{(2q+3)(2q+5)} \right)^2
 \; ,
\end{eqnarray}
$
 \beta= h /h_0(t)
$
is the {\em normalized halfwidth},
which depends on $t$,
and $h_0$ is given by (\ref{E9}).

%\subsection{Fixed Support in the Boundary Region}\label{V.1}

%We assume that $h\equiv t_0$ in the boundary region.
%All what we need to provide the continuity of the estimate is
%a continuous transformation of the boundary kernel
%such that at the touch point, $t_0$, the kernel turns out
%the interior kernel.

In the continuum limit, Eq.~(\ref{Bopt}) reduces to
$
 b_{q+2} = ((2q+3)z^2-1)\left/\left(
      \frac{2q+3}{(2q+5)\beta^{2q+5}} + \frac{2}{2q+5}
     \right)\right.
  .
$
Thus we have

\noindent {\bf Theorem~1.}
{\em
Among all boundary kernels with support $[0,2h]$,
the minimum leading order MSE is provided by the kernel
$K(t,x)=\frac{1}{h^{q+1}}G\left(\frac{t}{h}-1,\frac{x}{h}-1\right)$
where
\begin{equation}\label{E23}
G(z,y) \; = \;
 \gamma_q \cdot
 \left[
 P_q(y)
\; + \; (2q+3) z
 P_{q+1}(y)
\; + \;
 \frac{(2q+3)z^2-1}
{\frac{2q+3}{(2q+5)\beta^{2q+5}} + \frac{2}{2q+5}}
 P_{q+2}(y)
\right]
\; ,
\end{equation}
$P_q, P_{q+1}, P_{q+2}$ are the Legendre polynomials,
and $\beta=h/h_0(t)$.
}

We use $\beta =\frac{h}{h_0(t)}$ in Eq.~(\ref{E23}) in place of
$f^{(p)}(t)$ because we are interested in kernels which have a fixed
halfwidth in the boundary region:
$h(t)=h_0(t_0)$ and $\beta =\frac{h_0(t_0)}{h_0(t)}$
where $t_0$ is the touch point.

Let us examine the two special cases:
$t=h$ (when the estimation point is the touch point)
and $t=0$ (when the estimation point is the left edge of [0,1]).

When $t=h=h_0(t)$, the optimal boundary kernel is identical to
the optimal interior kernel (see (\ref{E119})).
Thus using the optimal boundary kernel
{\em guarantees the continuity of the estimate}
if at the touch point we apply the optimal interior kernel
of the optimal halfwidth.

In the edge case, $t=0$,
the halfwidth, $h_0(t)$ as given by Eq.\ (\ref{E9}),
is a singular point of the MSE functional
for the optimal boundary kernel.
In Appendix~B, we show that
the optimal boundary kernel with this halfwidth
has the minimum value of the leading order MSE
among all boundary kernels which have at most $p-1$ sign changes
in the open interval of their support (and thus are equivalent to
a local polynomial regression estimator with non-negative weighting;
see Section~\ref{LPR1}).
We name this kernel (with $t=0,\; h=h_0(0)$)
the {\em edge optimal kernel}.
The kernel has a simple expression:
$K(0,x)=\frac{1}{h^{q+1}} G\left(\frac{x}{h}-1\right)$
where $h=h_0(0)$ and
\[
G(y) \; = \;
 \gamma_q \cdot
 \left[
 P_q(y)
\; - \; (2q+3)
 P_{q+1}(y)
\; + \;
 (2q+2)
 P_{q+2}(y)
\right]
\; .
\]
%The kernel vanishes at the right end of the support, $x=2h$.
The leading order MSE equals
$
 4(q+1)^2
 \frac{2q+3}{2q+1}
 \frac{\sigma^2\gamma_q^2}{N h_0(t)^{2q+1}}
 ,
$
which is
{\em
exactly $4(q+1)^2$ times larger
than for the optimal interior kernel}.

The expressions for the optimal halfwidths
are identical for both interior and edge estimation cases.
This fact supports our suggestion to use a constant halfwidth
in the boundary region.

We assume that in the interior, $t\geq t_0$,
the optimal halfwidth, $h_0(t)$, is used.
Then the natural choice for the constant halfwidth in the boundary region
is $h=h_0(t_0)$:
this ensures the continuity of the estimate at the touch point.
When $N\rightarrow\infty$, the optimal halfwidth, $h_0(t_0)=t_0$
scales as $N^{-\frac{1}{2p+1}}$ and tends to zero.
Thus $\frac{f^{(p)}(\theta t_0)}{f^{(p)}(t_0)}\rightarrow 1$
and $\frac{h_0(\theta t_0)}{h_0(t_0)}\rightarrow 1$
uniformly for $\theta\in [0,1]$.

For $h(t)=h_0(t)$, the optimal boundary kernel
of Eq.~(\ref{E23})
is simplified to
\begin{equation}\label{E223}
G(z,y) \; = \;
 \gamma_q \cdot
 \left[
 P_q(y)
\; + \; (2q+3) z
 P_{q+1}(y)
 \; + \;
         ( (2q+3)z^2 -1 )
 P_{q+2}(y)
\right]
\; .
\end{equation}
Therefore,
we have

\noindent {\bf Corollary~2.}
{\em
If $h(t)\equiv h_0(t_0)$ in the boundary region,
the normalized optimal boundary kernel of Eq.~{\rm (\ref{E23})}
tends to
the kernel of Eq.~{\rm (\ref{E223})}
as $N\rightarrow\infty$.
}

We name this limit the {\em asymptotically optimal kernel}.
Its analytic expression is simpler and does not depend on $f^{(p)}(t)$,
but its leading order MSE might be larger than the optimal.
In contrast, the coefficient for $P_{q+2}$ in (\ref{E23}) depends on
$\beta$ which involves $f^{(p)}(t)$.
Section~\ref{Comp} shows that the asymptotically optimal kernel
achieves nearly the same MSE as the optimal kernel
when the kernel halfwidth is close to its optimal value.
As $N$ tends to infinity, the optimal halfwidth
can be estimated with increasing accuracy,
and little performance degradation results from using
the asymptotically optimal kernel instead of the optimal kernel.

Both the optimal and asymptotically optimal boundary kernels
are linear combinations of the low order Legendre polynomials.
In these linear combinations, only the coefficients $b_j$ depend on $t$.
When the interval of the support, $[0,2h]$, is the same for all
$t\in [0,h]$,
the kernel estimate is just a linear combination
of these $b_j(t)$ functions:
\[
 \widehat{f^{(q)}}(t) \; = \;
  \frac{\gamma_q}{h^{q+1}} \sum_j b_j \left(\frac{t-h}{h}\right)
         \left[
                 \int_0^{2h} P_j \left(\frac{x-h}{h}\right) Y(x) dx
         \right]
 \;
 ,
\]
where $Y(x)$ is the data.
In particular, the asymptotically optimal boundary kernel
produces an estimate, $\widehat{f^{(q)}}(t)$,
which is a quadratic function of $t$.

\section{Equivalence of Local Polynomial Regression and Kernel Estimators}\label{LPR1}

In the local polynomial regression, we minimize
\[
 F(a_0,a_1,\ldots,a_{p-1}) \; = \;
 \sum_{i=1}^N w_i(t) \cdot
 \left(\sum_{j=0}^{p-1} a_j(x_i-t)^j - y_i \right)^2
\]
and take $q!a_q$ as the estimate of $f^{(q)}(t)$.
The weights, $w_i(t)$, are non-negative and considered as given.
Since the functional is quadratic and non-negative,
the minimum exists and satisfies
\[
 0 \; = \; \frac{\partial F}{\partial a_k} \; = \;
 \sum_{j=0}^{p-1} \left[ \sum_{i=1}^N (x_i-t)^{k+j} w_i(t) \right] a_j
  \; - \; \sum_{i=1}^N (x_i-t)^k w_i(t) y_i
\]
for $k=0,1,\ldots ,p-1$.
This system of linear equations can be rewritten as
\begin{equation}\label{E59}
%\[
 \sum_{j=0}^{p-1} d_{kj}(t) \left( a_j h^j\right) \; = \; m_k(t) \; ,
 \;\;\;\;\;\;\;\; k=0,1,\ldots ,p-1 \; ,
%\]
\end{equation}
where
\[
 d_{kj}(t) \; = \; \frac{1}{Nh}
   \sum_{i=1}^N \left(\frac{x_i-t}{h}\right)^{k+j} w_i(t)
 \; , \;\;\;\;\;\;\;\;\;\;\;\;\;\;
 m_k(t) \; = \; \frac{1}{Nh}
   \sum_{i=1}^N \left(\frac{x_i-t}{h}\right)^k w_i(t) y_i
 \; .
\]
If the number of data points with non-zero weights is at least $p$,
the matrix $[d_{kj}(t)]$ is non-singular.
Let $[\tilde{d}_{jk}(t)]$ be the inverse matrix.
Then
$a_q h^q = \sum_{k=0}^{p-1} \tilde{d}_{qk}(t) m_k(t)$
and
$q! a_q = \sum_{i=1}^N K(t,x_i) y_i$,
where
\begin{equation}\label{E60}
 K(t,x_i) \; = \;  w_i(t)
                   \left[ \frac{q!}{Nh^{q+1}}
    \sum_{k=0}^{p-1} \tilde{d}_{qk}(t) \left(\frac{x_i-t}{h}\right)^k
                   \right]
 \; .
\end{equation}
%Therefore the equivalent kernel is the product of the weight function with
%a polynomial in $\frac{x-t}{h}$
%of order $p-1$
%whose coefficients might depend on $t$.
%One may easily verify that the equivalent kernel automatically satisfies
%the moment conditions and thus is a kernel of type $(q,p)$.
Thus
%we have
%
%\noindent {\bf Theorem~10.}
{\em
%For a given estimation point $t$ and weights $w_i$,
for a given estimation point $t$ and weights $w_i$,
the local polynomial regression estimator
is equivalent to a kernel estimator whose kernel
is the product of the weights with
a polynomial in $\frac{x_i-t}{h}$ of order $p-1$.
}
%One may easily check that
%the equivalent kernel automatically satisfies
The equivalent kernel automatically satisfies
the moment conditions and thus is a kernel of type $(q,p)$.
%}

We name the polynomial
in $x_i$
inside the brackets on the right hand side of (\ref{E60})
the {\em factor polynomial}.
We say that a discrete function $Q(x_i)$ has a sign change between
$x_j$ and $x_{j+k}$ if $Q(x_j)Q(x_{j+k})<0$
and $Q(x_{j+1})=\ldots =Q(x_{j+k-1})=0$.
The weights, $w_i(t)$, are non-negative,
and the factor polynomial has at most $p-1$ roots.
Therefore, for the given $t$,
the equivalent kernel $K(t,x_i)$ has at most $p-1$ sign changes.
Answering the question: ``which kernel estimators can be represented
as a local polynomial regression?''
we show that the necessary condition is also sufficient.

\noindent {\bf Theorem~3.}
{\em
A kernel of type $(q,p)$ is the equivalent kernel of
local polynomial regression of order $p-1$ with non-negative weights
if and only if the kernel has no more than $p-1$ sign changes.
}

It is known (see M\"uller (1985)) that any kernel of type $(q,p)$
has at least $p-2$ sign changes. This implies

\noindent {\bf Corollary~4.}
{\em
The actual order of the factor polynomial is at least $p-2$.
}

To solve system (\ref{E59}), we can expand
$w_i(t) \left(\frac{x_i-t}{h}\right)^{k+j}$
in orthogonal polynomials.
The representation of the equivalent kernel in terms of these polynomials
was described in Section~\ref{IV}
and corresponds to a QR-decomposition of the matrix $[d_{kj}]$.
Furthermore,
when the regression support in the boundary region is fixed,
one of the parts of this decomposition
(namely, the matrix of coefficients of the orthogonal polynomials)
is independent of the estimation point, $t$.
Thus the equivalent orthogonal polynomial representation
is a computationally convenient implementation
of the local polynomial regression in the boundary region.

\section{Optimal Weighting in Local Polynomial Regression}\label{LPR2}

It is known (M\"uller (1987), Fan(1993))
that the optimal interior kernel
of type $(q,p)$,
$p-q\equiv 0\,{\rm mod}\, 2$,
in the continuum limit,
is produced by the scaling weight function $W(y)=1-y^2$.
We show that this choice is not unique.
%Actually, different weightings may have the same
%equivalent kernel and thus may produce the same estimator.

\noindent {\bf Theorem~5.}
{\em
Let $p-q$ be even.
If data points, $x_i$, in the interval of support, $[t-h,t+h]$,
are symmetric around the estimation point, $t$,
and their weights are chosen as $w_i=W\left(\frac{x_i-t}{h}\right)$,
then each of the functions
$W_1(y)=1-y$, $W_2(y)=1+y$, $W_3(y)=1-y^2$
produces the same estimator.
}

Because of the optimality in the interior,
the Bartlett-Priestley weighting, $W(y)=1-y^2$,
is used often in the boundary region as well
(Hastie and Loader (1993)).
Since the kernel support is not symmetric
around the estimation point,
choosing
the Bartlett-Priestley weighting
is somewhat arbitrary.
Even in the limiting case, this weighting does not
provide the minimum MSE.

As we showed in Section~\ref{IV},
the leading order MSE is minimal when we use the optimal kernel
given by Eq.~(\ref{E4}) and~(\ref{Bopt}).
In the interval of its support, this is a polynomial of order $p$.
If the optimal kernel has no more than $p-1$ sign changes,
there exists an equivalent weighting in the local polynomial regression.
The true order of the factor polynomial is either $p-1$ or $p-2$.
Then the optimal weighting, which provides
the minimum value of the leading order MSE,
%in local polynomial regression,
must be a linear or quadratic function.

The optimal boundary kernel depends on the derivative
$f^{(p)}(t)$ whose value might be unknown.
This dependence is eliminated in the asymptotically optimal
kernel which approximates the optimal kernel as $N\rightarrow\infty$.
For the case $p=q+2$, the asymptotically optimal kernel
was determined in Section~{V}, Eq.~(\ref{E223}).

The following result shows that the asymptotically optimal kernel
is representable as a local polynomial regression estimator
with a non-negative weight function.

\noindent {\bf Theorem~6.}
{\em
The asymptotically optimal kernel
has no more than $p-1$ roots in the open interval of its support,
$[0,2h]$.
Its equivalent weighting is a linear function
which is non-negative on $[0,2h]$.
In the case of edge estimation $(t=0)$,
the equivalent weighting equals $2h-x$.
For the touch point $(t=t_0)$,
the equivalent weighting
can be chosen as either $2h-x$ or $x$
and produces the same estimate as the Bartlett-Priestley weighting.
}

For the intermediate estimation points, $0<t<t_0$,
{\em
the slope of the weighting line varies as $t$ changes.
}
For example, if $q=0$, the equivalent weighting
can be represented as
$(1-z^2)h+\left( z+\sqrt{1-3z^2+3z^4}\right) (x-h)$
where $z=\frac{t-h}{h}$.

In the case when we allow variable halfwidth in the boundary region,
we have to optimize simultaneously the kernel shape (or weighting)
and halfwidth.
We carry out this optimization in Appendix~B.
The optimal weighting in this case is
the linear function $2h(t)-x$ which
vanishes at the right end of the support.
(The vanishing guarantees
the continuity of the estimate as a function of $t$.)
The optimal halfwidth, $h=h(t)$, is a root of a polynomial equation.
The equivalent kernel
does not transform into the optimal interior kernel, however.
Thus this estimator is useful only if we estimate $f^{(q)}(t)$
in the boundary region and not in the interior.

\section{Estimation Near the Boundary for Discrete Data}\label{LPR3}

For discrete data, we propose two slightly different methods
for estimation near the boundary.
Both of them reduce MSE
relative to the local polynomial regression
with Bartlett-Priestley weighting.
First, the kernel of Section~\ref{IV}
is optimal for an arbitrary placement of points in the boundary region.
The kernel estimate can be computed
in ${\cal O}(N_T+N_E)$ operations,
where $N_E$ is the number of estimation points $t$
in the boundary region
(see Section~\ref{OK}).

The second method is to use the local polynomial regression
with the asymptotically optimal weighting
(the linear weighting function given in Theorem~6).

In the interior, when the data points are not equispaced,
the requirement that the kernel
vanishes at the ends of its support
(in order to ensure the continuity of the estimate)
is in conflict with the kernel shape optimization.
The equivalent kernel of
the local polynomial regression with Bartlett-Priestley weighting
vanishes at the ends of the support
and is asymptotically close to the optimal kernel.
Thus we agree with Hastie and Loader (1993) that
in the interior of the data interval,
the local polynomial regression
is the best way to estimate $f^{(q)}$.

To combine estimation in the interior and boundary region
we should make sure that they produce the same estimate
at the touch point.
In the continuum limit case, this holds (see Theorem~6)
because the equivalent kernel of the local polynomial regression
turns out the polynomial of order $p$.
This still holds in the equispaced data case if $t_0$ is one
of the data points or the midpoint of two data points.
In such a case, Theorem~5 guarantees that for $t=t_0$
the linear weighting $2h-x$ produces the same estimator
as the Bartlett-Priestley weighting, $h^2-(x-h)^2$.
For an arbitrary pattern of data points,
the equivalent polynomial of
the local polynomial regression with Bartlett-Priestley weighting
is generally of order $2+(p-1)=p+1$.
Thus, estimates
$\widehat{f}_{\rm boundary}^{(q)}(t_0)$
and
$\widehat{f}_{\rm interior}^{(q)}(t_0)$
will differ.
This discrepancy is eliminated by setting
$$
\widehat{f^{(q)}}(t)
  \; = \;
\widehat{f}_{\rm boundary}^{(q)}(t)
   \; - \;
 \frac{t}{t_0} \left[
\widehat{f}_{\rm boundary}^{(q)}(t_0)
  -
\widehat{f}_{\rm interior}^{(q)}(t_0)
   \right]
 .
$$
The correction term in the brackets
vanishes identically for equispaced data and
is asymptotically small as $N\rightarrow\infty$.

%Such a weighting is not symmetric around the estimation point,
%and thus was not considered in the literature before.

\section{Comparison of Different Estimators}\label{Comp}

We compare, in the continuum limit, the performance of our
%optimal, asymptotically optimal, 0-optimal and 1-optimal
boundary kernels
and local polynomial regression with Bartlett-Priestley weighting.
The latter estimator
is equivalent to the boundary kernel
$K(t,x)=\frac{1}{h}G\left(\frac{t}{h}-1,\frac{x}{h}-1\right)$
where
\[
G(z,y) \; = \;
 P_{q}(y)
\; + \; (2q+3) z
 P_{q+1}(y)
\; - \;
 (1 + (2q+3)z + b(z))
 P_{q+2}(y)
\; + \;
 b(z)
 P_{q+3}(y)
\; ,
\]
and
$P_{q}, P_{q+1}, P_{q+2}, P_{q+3}$ are the Legendre polynomials.
For the touch point, $z=0$, we have $b(0)=0$.
In particular, if $q=0$ then
$b(z)=\frac{9z^2}{10z^2-8z+1}$.
%We remind that our kernels are equivalent
%to some other weightings (linear or quadratic)
%in the local polynomial regression.

Figure~1 plots $R(t)/R(t_0)$
(the MSE at estimation point $t$ normalized to
the MSE at the touch point)
as a function of $t$
for the optimal boundary kernel when $h(t)=h_0(t)$.
When the estimation point approaches the edge,
MSE is $4(q+1)^2$ times larger than in the interior.
Figure~2 compares the ratio of the MSE
for the Bartlett-Priestley weighting of local polynomial regression
with the optimal kernel.
For kernels of type (0,2),
there is an improvement of at most five percent.
For type (4,6), the perfomance ratio increases to 21\% at its highest.
The difference is largest when $t>0.5$.
For smaller $t$, the half parabola of the Bartlett-Priestley weighting
resembles the equivalent linear weighting of the optimal kernel,
and thus the MSEs are similar.

Figures~1 and~2 are calculated assuming $h(t)=h_0(t)$.
We now consider the case where the kernel halfwidth is different
than the optimal halfwidth.
This case occurs when $|f^{(p)}|^2$ is estimated poorly
or when $h(t)$ is determined by other requirements.
Figure~3 plots the ratio of the MSE
of the Bartlett-Priestley weighting to that of our kernel
for $h(t)=\frac{1}{2}h_0(t)$ and $h(t)=2h_0(t)$.
We see that both kernels perform similarly when $h(t)$ is less
than $h_0(t)$. However, when $h(t)$ is greater than $h_0(t)$,
the optimal kernel performs much better than
the Bartlett-Priestley weighting.
This occurs because the optimal kernel has better bias protection.

Figure~4 gives the same plot for the asymptotically optimal kernel
(or, equivalently, for the asymptotically optimal linear weighting
in local polynomial regression).
The difference in performance is less because
the shape of the asymptotically optimal kernel is independent of
$h(t)/h_0(t)$.
For some values of $t$, the Bartlett-Priestley weighting
actually outperforms the asymptotically optimal weighting.
This occurs because the asymptotically optimal kernel is optimal
only when $h(t)=h_0(t)$.
Figure~4 shows that the performance of the two weightings is similar for
$h(t)\leq h_0(t)$. However, the asymptotically optimal weighting
has an appreciable advantage over the Bartlett-Priestley weighting
for $h(t)>h_0(t)$.

M\"uller (1991), M\"uller and Wang (1994)
suggest boundary modifications
of the optimal interior kernels.
Their modifications are done under the constraint that the kernel
vanishes at both endpoints of its support.
%In the interior, such a requirement is natural to provide
%smoothness of the estimate.
We agree with Hastie and Loader (1993, p.140) and Jones (1994, p.10)
that this requirement is artificial in the boundary region.
The M\"uller boundary kernel of type $(q,q+2)$
is the unique polynomial of order $q+3$ which satisfies
simultaneously
the moment conditions and the two boundary conditions:
$K(t,0)=0$ and $K(t,2h)=0$.
This kernel is
$K(t,x)=\frac{1}{h^{q+1}}G\left(\frac{t}{h}-1,\frac{x}{h}-1\right)$
where
%\begin{equation}\label{E163}
\[
G(z,y) \; = \;
 \gamma_q \cdot
 \left[
 P_q(y)
\; + \; (2q+3) z
 P_{q+1}(y)
\; - \;
 P_{q+2}(y)
\; - \; (2q+3) z
 P_{q+3}(y)
\right]
\; .
\]
%\end{equation}
%The restriction $K(t,0)=0$ is useful when data
%is being added in time.
%In this case, the estimate using the M\"uller kernel depends
%less sensitively on the recent data
%than using the kernels of Sections~\ref{V.2} and~\ref{V.3}.
The requirement that the kernel vanishes at the left end point
leads to a significantly larger MSE
(especially when we are estimating close to the edge).
The comparison of MSE for the M\"uller kernel
and for our optimal boundary kernel
is given on Figure~5.
%For $t<0.5$,
The M\"uller kernel has noticably larger risk.

\section{Conclusion}

In Section~\ref{IV}, we have derived a smoothing kernel which
minimizes the leading order expected mean square error
for a given pattern of data points $x_i$.
In Section~\ref{OK}, we have described
a fast algorithm to compute the estimate
in the case when the kernel halfwidth is constant
in the boundary region.
In Section~\ref{V}, we have found an explicit formula
for the optimal boundary kernel of type $(q,q+2)$
in the continuum limit
(when the data points are spaced approximately regularly
and their number tends to infinity).
We also have defined the asymptotically optimal boundary kernel
which is an approximation of the optimal kernel
where dependence on
$f^{(p)}(t)$
is eliminated.
Both kernels are polynomials of order $q+2$
whose coefficients depend on $t$.
When the estimation point is the first or the last in the dataset,
the minimal possible MSE is $4(q+1)^2$ larger in comparison with
the estimation in the interior.
These boundary kernels can also be used for prediction
with a minimum of MSE.
In Section~\ref{LPR1},
we have proved that a kernel estimator of type $(q,p)$
is equivalent to a local polynomial regression estimator of order $p-1$
with some non-negative weighting
if and only if the kernel has at most $p-1$ sign changes in its support.
In Section~\ref{LPR2}, we have shown that
the asymptotically optimal boundary kernel
of type $(q,q+2)$
is equivalent to a local polynomial regression with
non-negative linear weighting
whose slope depends on the estimation point.
In Section~\ref{LPR3}, we have described how to apply
the optimal boundary kernels and weightings
to discrete data with arbitrarily placed points.
In Section~\ref{Comp}, we compare MSE of our kernel estimators
versus local polynomial regression with
the Bartlett-Priestley weighting.
The optimal boundary kernel takes into account changes in
$f^{(p)}(t)$ and thus always outperforms the local polynomial
regression.
The asymptotically optimal boundary kernel
is more robust than the local polynomial regression
with respect to misspecification of the halfwidth
(caused by possible errors in estimating $f^{(p)}$).

\vspace{8mm}

\noindent
{\Large
{\bf Appendix~A.\/\/ Legendre Polynomials}
}

\vspace{3mm}

Let $P_0,P_1,\ldots\;$ be the Legendre
polynomials on $[-1,1]$ :
\[ P_0(y)=1 \; , \;\;\;\; P_1(y)=y \; , \;\;\;\;
   P_i(y) \; = \; \frac{1}{i}[(2i-1)yP_{i-1}(y) - (i-1)P_{i-2}(y)] \; . \]
Set
\[ m_{ij} \; = \; \int_{-1}^1 P_i(y)\frac{1}{j!}y^j dy \;
  = \; \left\{ \begin{array}{l}
       \frac{2^{i+1}\left(\frac{i+j}{2}\right)!}
            {(i+j+1)!\left(\frac{j-i}{2}\right)!}
       \;\;\;\; {\rm if} \;\; j\geq i, \; j\equiv i\,{\rm mod}\, 2\;\; ;
                  \\
        0 \; , \;\;\;\; {\rm otherwise} \; .
             \end{array} \right. \]
In particular,
$
 m_{q,q}  =  \frac{1}{\gamma_q}
$ and $
 m_{q-1,q+1}  =  \frac{1}{2\gamma_q}  ,
$
where
$ \gamma_q  =  \frac{1}{2} \prod_{i=1}^{q}(2i+1)  . $

For the interval $[-h_L,h_R]$, we define
$
  \overline{P}_i(x)   =   P_i\left( 2\frac{h_L+x}{h_L+h_R}-1\right)
            =   P_i\left( z+\frac{x}{h}\right)
 ,
$
where
$
  h=\frac{1}{2}(h_L+h_R) , \;
   z=\frac{h_L}{h}-1=\frac{h_L-h_R}{h_L+h_R}
 .
$
Then
\[ \int_{-h_L}^{h_R} \overline{P}_i(x)\overline{P}_j(x)dx \; = \; h\int_{-1}^1 P_i(y)P_j(y)dy
   \; = \; \frac{2h}{2i+1} \delta_{ij} \; . \]
Define
\[ C_{ij} \; = \;
 \left(\frac{1}{h}\right)^{j+1} \int_{-h_L}^{h_R} \overline{P}_i(x)\frac{x^j}{j!} dx
 \; = \;
%    \int_{-h_L}^{h_R} P_i\left( \frac{x}{h}+z\right) \frac{1}{j!}
%    \left( \frac{x}{h}\right)^j \frac{dx}{h} \; = \]
%\[ = \;
%    \int_{-1}^1 P_i(y)\frac{(y-z)^j}{j!} dy
%     = \;
   \sum_{k=0}^{j}
 \frac{(-1)^{j-k} z^{j-k}}{(j-k)!}
                       \int_{-1}^1 P_i(y)\frac{y^k}{k!} dy \; . \]
Since $m_{ik}=0$ with $i>k$, we have
$
C_{ij}=0
$
if $i>j$,
and
$
C_{ij}=
\sum_{k=i}^{j}\frac{(-1)^{j-k} z^{j-k}}{(j-k)!} m_{ik}
$
if $i\leq j$.
For the case of $p =q+2$, this gives
 \[ C_{qq} \; = \; m_{qq} \; ,
           \;\;\;\;\;\;
    C_{q,q+1} \; = \; - z m_{q,q} \; ,
           \;\;\;\;\;\;
    C_{q,q+2} \; = \; m_{q,q+2} + \frac{1}{2} z^2 m_{q,q} \; , \]
 \[ C_{q+1,q+2} \; = \; - z m_{q+1,q+1} \; ,
           \;\;\;\;\;\;
   C_{q+1,q+1} \; = \; m_{q+1,q+1} \; ,
           \;\;\;\;\;\;
   C_{q+2,q+2} \; = \; m_{q+2,q+2} \; . \]
Equation (\ref{4.2}) reduces to
$
 b_q  =  \frac{1}{C_{qq}}  =  \gamma_q
$ and $
 b_{q+1}  =  - \frac{1}{C_{q+1,q+1}}C_{q,q+1}b_q  =
  (2q+3) z  \gamma_q  .
$

\vspace{8mm}
\newpage
\noindent
{\Large
{\bf Appendix~B.\/\/
Simultaneous Optimization of

\hspace{33mm}
the Kernel Shape and Halfwidth
}
}

\vspace{3mm}

Granovsky and M\"uller (1989)
derived the optimal shape of interior kernel of type $(q,p)$
as a function which minimizes the leading order MSE and has
$p-2$ sign changes in the interval of its support.
When the support is not symmetric around the estimation point,
$p-2$ sign changes are not enough to meet the moment conditions.

Thus for boundary kernels we have to relax the limitation
and allow $p-1$ sign changes.
By Theorem~3, kernels that fit this limitation
are equivalent kernels of the local polynomial regression estimators.
Therefore, optimization of the kernel shape and support,
subject to this limitation,
is equivalently optimization of the weighting function
for the local polynomial regression
in the case when the halfwidth is not fixed.

\noindent {\bf Theorem~7.}
{\em
Let $t$ be an estimation point in the left boundary region,
and consider the class of boundary kernels of type $(q,p)$
with at most $p-1$ sign changes in their support.
In the continuum limit,
the leading order MSE is minimized
when the kernel is a polynomial of order $p$
within its support interval
and vanishes at the right end of its support.
}

The proof of Theorem~7 resembles the proof
of the main theorem of Granovsky and M\"uller (1989).
The only difference is that they considered kernels as functions
from $L_2(-\infty ,\infty )$ while we need $L_2[-t,\infty )$.
Their proof shows that the optimal kernel is a continuous function,
has finite support, and is a polynomial of order $p$ there.
In the case of space $L_2[-t,\infty )$,
it implies that the optimal kernel vanishes
at the right end of its support.

The kernel defined in Theorem~7 is unique.
Indeed, in its Legendre polynomials expansion,
$
 K(t,x)
=
\frac{1}{h(t)^{q+1}}
G\left(\frac{t}{h(t)}-1,\frac{x}{h(t)}-1\right)
 ,
 \;
$
$
G(z,y)
=
\sum_{k=q}^p b_k(z) P_k(y)
,
$
coefficients $b_q,b_{q+1},\ldots ,b_{p-1}$ are determined from
the moment conditions,
and the last coefficient is fixed by the requirement to vanish
at the right end:
$b_p = - (b_q+b_{q+1}+\ldots +b_{p-1})$.
The leading order MSE for this kernel is a rational function in $h$.
Thus the optimal halfwidth is a root of a polynomial equation
and depends on $t$ and $|f^{(p)}(t)|^2$.
For instance, in the case $p=q+2$,
$$
G(z,y) \; = \;
 \gamma_q \cdot
 \left[
 P_q(y)
\; + \; (2q+3) z
 P_{q+1}(y)
 \; - \; ( 1 + (2q+3)z )
 P_{q+2}(y)
\right]
\; ,
%\eqno{({\rm B}.1)}
$$
and the optimal halfwidth equals
$h=\beta\cdot h_0(t)$, where $\beta$
is the maximal root of
$$
 (2q+2)\beta^{2q+6} - (4q+8)\tau\beta^{2q+5} + (2q+5)\tau^2\beta^{2q+4}
 - (2q+2)\beta + (2q+3)\tau
 \; = \; 0 \; ,
$$
with $\tau =t/h_0(t)$.
In the edge estimation case, $t=0$, we have $\beta =1$,
and therefore the optimum is attained for the halfwidth
$h=h_0(0)$.
This implies that our edge optimal kernel attains the minimum MSE
among all boundary kernels of type (q,q+2) which have at most
$q+1$ sign changes in the support.

For all $\tau$, we have $\beta >\tau$,
and the optimal halfwidth, $h(t)$, is always larger than $t$,
so there is no touch point.
For any $q,p$,
the optimal boundary kernel of Theorem~7 always has
a non-symmetric support and always differs from
the optimal interior kernel.
This result is natural because the latter has fewer sign changes.

The fact that the optimal kernel of Theorem~7 vanishes
at the right end of the support guarantees
the continuity of the estimate as a function of $t$.

The local polynomial regression with the linear weighting
$2h-x,\;$
$x\in [0,2h]$,
is equivalent to the optimal kernel of Theorem~7.
Indeed, the equivalent kernel is a polynomial of order $1+(p-1)=p$
and vanishes at the right end of the support.
Thus we have

{\bf Corollary~8.}
{\em
In the case when the halfwidth is not fixed,
the leading order MSE of
local polynomial regression
is minimized for the linear weighting $2h-x$.
The optimal halfwidth, $h=h(t)$, depends on $|f^{(p)}(t)|^2$.
}

\vspace{8mm}

\noindent
{\Large
{\bf Appendix~C.\/\/ Proofs of Theorems~3, 5, 6}
}

\vspace{3mm}

{\bf Lemma~9.}
{\em
Let $K_1(x_i)$ and $K_2(x_i)$ be kernels of type $(q,p)$
with the same estimation point and the same support
such that $K_r(x_i)=W(x_i)Q_r(x_i)\;$, $r=1,2$,
where $W(x_i)\geq 0$ for all data points $x_i$ in the support.
If $Q_1(x)$ and $Q_2(x)$ are polynomials of order $p-1$
then $K_1(x_i)=K_2(x_i)$ for every data point $x_i$.
}

\noindent
{\bf Proof.}
Since $K_1$ and $K_2$ satisfy the same moment conditions,
their difference is orthogonal to any polynomial $P(x_i)$
of order $p-1:\;$
$
 \sum_i (K_1(x_i)-K_2(x_i))P(x_i) = 0
 .
$
When we choose
$P(x_i)=Q_1(x_i)-Q_2(x_i)$,
we have
$
 \sum_i W(x_i) (Q_1(x_i)-Q_2(x_i))^2 = 0
 .
$
Since $W(x_i)\geq 0$, it implies
$
 W(x_i) (Q_1(x_i)-Q_2(x_i)) = 0
$
for every $x_i$.

{\bf Proof of Theorem~3.}
Let a kernel $K(x_i)$ have $m\leq p-1$ sign changes
in the interval of its support.
We enumerate the sign changes: $z_1,z_2,\ldots ,z_m$.
Namely, if the $l$th sign change occurs
at $x_j$ or between $x_j$ and $x_{j+k}$,
we set $z_l=x_j+\varepsilon$ where
$\varepsilon < \min\{ x_2-x_1,x_3-x_2,\ldots ,x_N-x_{N-1}\}$.
Now we define $P(x)=(-1)^s \prod_{l=1}^m (x-z_l),\:$
$W(x_i)=K(x_i)/P(x_i)$.
The function $W(x_i)$ has no sign changes.
We choose $s$ to make all of the values $W(x_i)$ non-negative.
Let $Q$ be the factor polynomial for
the local polynomial regression with the weights $w_i=W(x_i)$.
Since $K=WP$ and $WQ$ are kernels of type $(q,p)$,
and $P,Q$ are polynomials of order $p-1$,
Lemma~9 implies that
$K(x_i)=W(x_i)P(x_i)=W(x_i)Q(x_i)$
for every data point $x_i$.
Thus $K$ is the equivalent kernel for
the local polynomial regression with the weights $w_i$.

{\bf Proof of Theorem~5.}
It is sufficient to check that weightings $W_1(y)=1-y$ and $W_3(y)=1-y^2$
have the same equivalent kernel.
Let $Q_1(y)$ and $Q_3(y)$ be their respective factor polynomials.
Since $Q_3$ is a polynomial of order $p-1$,
then $W_3Q_3$ is a polynomial of order $p+1$.
Since $W_3$ is even
and the placement of data points is symmetric,
the equivalent kernel, $W_3Q_3$, is an even function
(if $q$ is even) or an odd function (if $q$ is odd).
The difference $p-q$ is even, thus $W_3Q_3$
can not have term $y^{p+1}$.
Therefore, $W_3Q_3$ is a polynomial of order $p$,
and the true order of $Q_3$ is at most $p-2$.
Now we notice that $W_3(y)Q_3(y)=W_1(y)\left[ (1+y)Q_3(y)\right]$.
Both $(1+y)Q_3(y)$ and $Q_1(y)$ are polynomials of order $p-1$.
Thus Lemma~9 implies that $W_3(y)Q_3(y)=W_1(y)Q_1(y)$
when $y=\frac{x_i-t}{h}$.

{\bf Proof of Theorem~6.}
First, we show that for every $-1\leq z\leq 0$,
the normalized kernel $G(y)=G(z,y)$, given by Eq.\ (\ref{E23}),
has at least one root outside (-1,1).
Indeed, since $P_k(1)=1$, $P_k(-1)=(-1)^k$,
we have
$G(1)=\gamma_q (2q+3)z(z+1)$,
$G(-1)=\gamma_q (2q+3)z(z-1)$.
If $z=0$, we have $G(1)=G(-1)=0$.
If $z=-1$, we have $G(1)=0$.
If $z\neq 0,\: z\neq -1$, we have
$sign\, G(1)\cdot sign\, G(-1)=(-1)^{q+1}$.
On the other hand,
$G(y)/G(-y) \rightarrow (-1)^q$ as $y\rightarrow\infty$.
Therefore, $G$ must have a root
$y_0=y_0(z)$
either in $[1,\infty )$ or in $(-\infty ,-1]$.
Since $G$ is a polynomial of order $p$, the number of roots within (-1,1)
is at most $p-1$.
Representing $G(y)=|y-y_0|Q(y)$
and applying the continuous version of Lemma~9,
we conclude that the local polynomial regression
with the linear weighting $|y-y_0|$,
estimation point $z$,
and support [-1,1],
has $G(y)$ as its equivalent kernel.
If $z=-1$ (the edge estimation case),
then $G(1)=0$, $y_0=1$, and the equivalent linear weighting is $1-y$.
If $z=0$ (the touch point estimation),
then $G(1)=G(-1)=0$, and either of $1-y$ and $1+y$ weightings
has $G$ as the equivalent kernel.
By the continuous version of Lemma~9,
the weighting $1-y^2$ also has $G$ as its equivalent kernel.

\vspace{8mm}

\noindent
{\bf Acknowledgements}

We thank C. Hurvich and A. Mullhaupt for useful discussions.
The referees helpful comments are gratefully acknowledged.

\newpage

\noindent {\bf References}

%Altman, N. S.
%(1990). {Kernel smoothing with correlated errors.}
%{\em J. Amer.\ Statist.\ Assoc.} {\bf 85} 749-759.

Cleveland, W. S. (1979).
Robust locally weighted regression and smoothimg scatterplots.
{\em J. Amer.\ Statist.\ Assoc.} {\bf 74} 829-836.

Fan, J. and Gijbels, I. (1992).
Variable bandwidth and local linear regression smoothers.
{\em Ann.\ Statist.} {\bf 20} 2008-2036.

Fan, J. (1993).
Local linear regression smoothers and their minimax efficiencies.
{\em Ann.\ Statist.} {\bf 21} 196-216.

Gasser, T. and M\"uller, H. G. (1979).
Kernel estimation of regression functions.
In: {\em Lecture Notes in Mathematics}, {\bf 757}, pp.23-68.
New York: Springer.

Gasser, T., M\"uller, H. G. and Mammitzsch, V. (1985).
Kernels for nonparametric curve estimation.
{\em J. Roy.\ Statist.\ Soc.\ Ser.\ B} {\bf 47} 238-252.

Granovsky, B. L. and M\"uller, H. G. (1989).
On the optimality of a class of polynomial kernel functions.
{\em Statistics and Decisions} {\bf 7} 301-312.

Grenander, U. and Rosenblatt, M. (1957).
{\em Statistical analysis of stationary time series.}
New York: Wiley.

%Hardle, W., Hall, P., and Marron, S.
%(1988). {How far are automatically chosen smoothing
%parameters from their optimum?}
%{\em J. Amer.\ Statist.\ Assoc.} {\bf 83} 86-95.

Hardle, W. (1990). {\em Applied nonparametric regression.}
Cambridge, New York:
Cambridge University Press.

Hastie, T. and Loader, C. (1993).
Local regression: automatic kernel carpentry.
{\em Statistical Science} {\bf 8} 120-143.

Jones, M. S. (1994).
Simple boundary correction for kernel density estimation.
{\em Statistics and Computing}, to appear.

Lejeune, M. (1985).
Estimation non-param\'etrique par noyaux:
regression polynomial mobile.
{\em Revue de Statistiques Appliqu\'ees} {\bf 33} 43-67.

M\"uller, H. G. (1984).
Smooth optimum kernel estimators of regression curves, densities and modes.
{\em Ann.\ Statist.} {\bf 12} 766-774.

M\"uller, H. G. (1985).
On the number of sign changes of a real function.
{\em Periodica Mathematica Hungarica} {\bf 16} 209-213.

M\"uller, H. G. (1987).
Weighted local regression and kernel methods
for nonparametric curve fitting.
{\em J. Amer.\ Statist.\ Assoc.} {\bf 82} 231-238.

M\"uller, H. G. (1988).
{\em Nonparametric Regression Analysis of Longitudinal Data}.
New York: Springer.

M\"uller, H. G. (1991).
Smooth kernel estimators near endpoints.
{\em Biometrika} {\bf 78} 521-530.

M\"uller, H. G. and Stadtm\"uller, U. (1987).
Variable bandwidth kernel estimators of regression functions.
{\em Ann.\ Statist.} {\bf 15} 610-625.

M\"uller, H. G. and Wang, J. L. (1994).
Hazard rate estimation under random censoring with varying
kernels and bandwidths.
{\em Biometrics}, to appear.

Rice, J. (1984).
Boundary modification for kernel regression.
{\em Comm. Statist.} A {\bf 13} 893-900.

Rice, J. and Rosenblatt, M. (1981).
Integrated mean square error of a smoothing spline.
{\em J. Approx. Theory} {\bf 33} 353-369.

Silverman, B. W. (1984).
Spline smoothing: the equivalent variable kernel method.
{\em Ann.\ Statist.} {\bf 12} 898-916.

{Stone, C. J. (1982).
{Optimal global rates of convergence for nonparametric regression.}
{\em Ann.\ Statist.} {\bf 10} 1040-1053.}

%\end{document}

%\emem{ Hall, P., Sheather, S.J., Jones, M.C. and Marron, J.S.
%(1991). {On optimal data-based bandwidth estimation.}
%{\em Biometrika} {\bf 78}, 263-269. }

%\noindent{ Hart, D., and Wehrly, T.E.
%(1986). {Kernel regression estimation using repeated measurments data.}
%{\em J. Amer.\ Stat. Assoc.} {\bf 81}, 1080-1088. }

%\noindent{ M\"uller, H.G.
%(1980). {\em Nonparametric regression analysis of longitudinal data.}
%Springer Verlag,  Berlin, New York.}

%\noindent{Priestley, M.B. (1981).
%{\em Spectral analysis and timeseries.} Ch. 11, { Academic Press}, New York. }

%\noindent{Riedel, K.S.  (1992).
%Data-based determination of evolutionary spectra.
%Accepted for publication in I.E.E.E. Trans. in Signal Processing} 

%\noindent{Wahba, G. (1980). {Automatic smoothing of the log periodogram.}
%{\em J. Amer.\ Stat. Assoc.} {\bf 75}, 122-132. }

\vspace{8mm}

%\begin{center}
%Alexander Sidorenko  \/ \/ \/
%and  \/ \/ \/ Kurt S. Riedel \\
%Courant Institute of Mathematical Sciences \\
%New York University \\
%New York, New York 10012
%\end{center}
%[pages=-,pagecommand={},width=\textwidth]
\includepdf[pages=-,pagecommand={}]{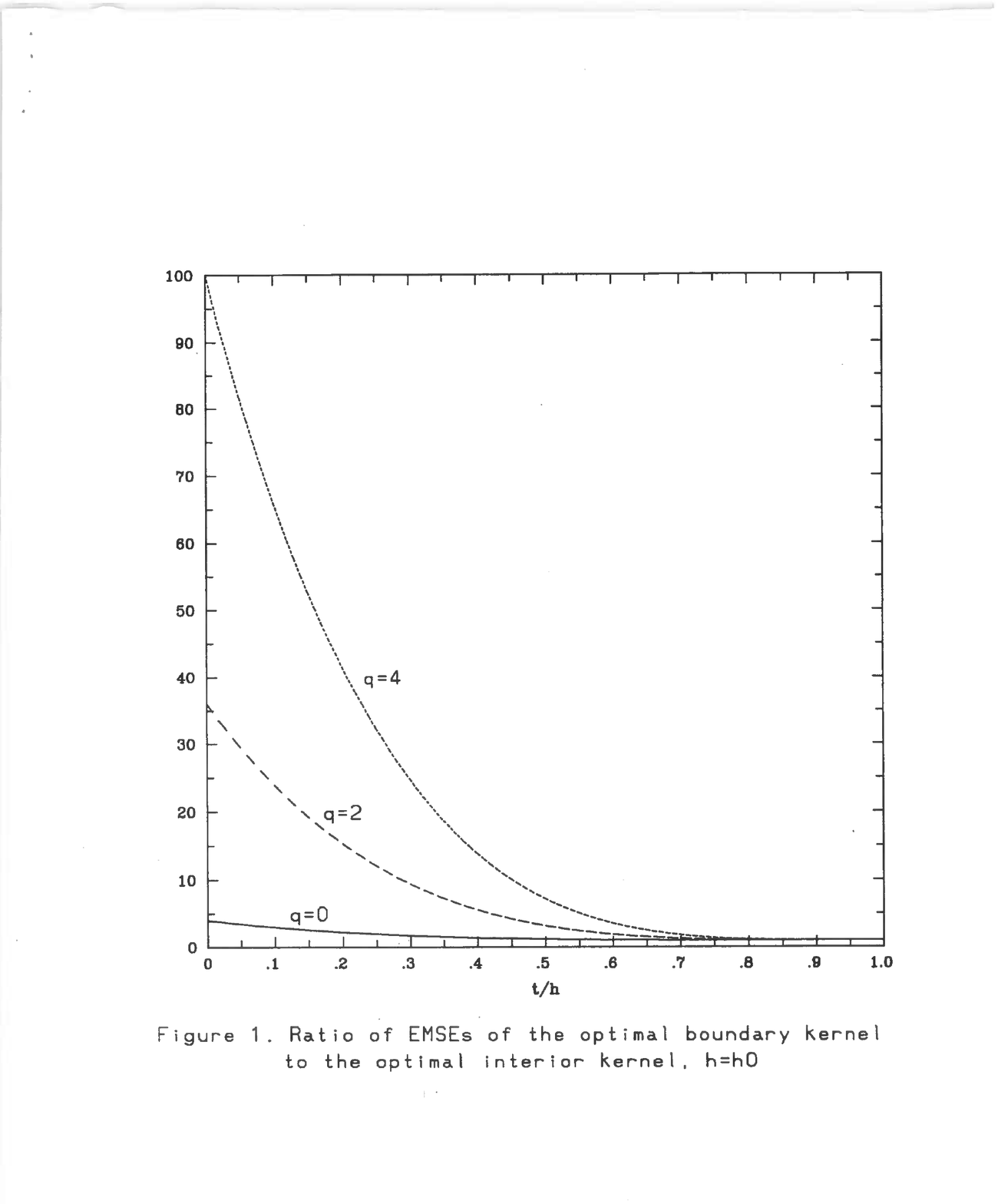}
\end{document}